# Observation of Ferromagnetic Clusters in $Bi_{0.125}Ca_{0.875}MnO_3$

Yuhai Qin and Trevor A. Tyson

*New Jersey Institute of Technology, Newark, NJ 07102, USA*

Klaus Pranzas and Helmut Eckerlebe

*GKSS Research Center,D-21502 Geesthacht, Germany*

## Abstract

The electron doped manganite system, $Bi_{0.125}Ca_{0.875}MnO_3$, exhibits large bulk magnetization of unknown origin. To select amongst possible magnetic ordering models, we have conducted temperature and magnetic field dependent small-angle neutron scattering measurements. Nontrivial spin structure has been revealed. Ferromagnetic spin clusters form in the antiferromagnetic background when temperature is decreased to $T_c \sim 108K$. With a further reduction in temperature or the application of external magnetic field, the clusters begin to form in larger numbers, which gives an overall enhancement of magnetization below $T_c$.

# I. INTRODUCTION

The colossal magnetoresistance (CMR) effect in perovskite manganite, $A_{1-x}B_xMnO_3$ (where A is a tri-valent rare-earth cation and B is a divalent cation), has attracted much interest in the past decade. A complete understanding of CMR depends on the knowledge of competing magnetic, structural and charge interactions in these compounds. Among the oxide systems investigated, $Bi_{1-x}Ca_xMnO_3$ constitutes a very interesting but less-studied system. Previous work ([1~4]) have revealed that the low-temperature resistivity drops continuously when $CaMnO_3$ is doped with $Bi^{3+}$ and reaches a minimum resistivity at x~0.875 with a weak temperature dependence. At x~0.875, a sharp increase in magnetization with decreasing temperature was found, which suggests an anomalous magnetic transition.

Tyson *et al.* have measured magnetization vs. external field (pulsed) [2, 4]. The result (Fig.1) shows that even in very high magnetic fields (up to 60T), the magnetization never saturates (and never approaches the theoretical limit of 3.1 Bohr magnetons per Mn site), and implies a strongly correlated spin configuration in this system.

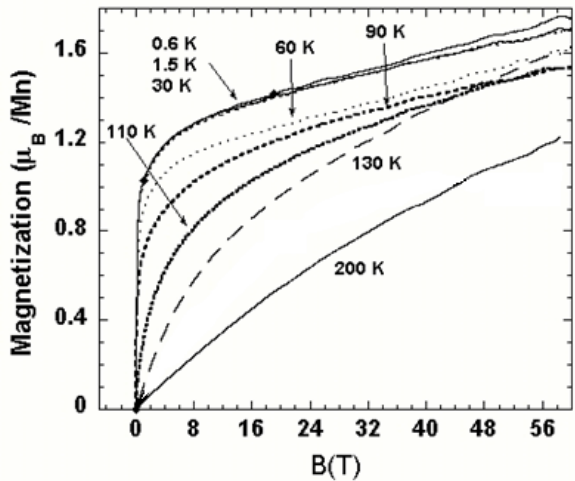

**Fig.1** Magnetization vs. external field for $Bi_{0.125}Ca_{0.875}MnO_3$.[2]

Historically, there are two interpretations concerning the ferromagnetic component in similar systems (like $Pr_{1-x}Ca_xMnO_3$). Jirak *et al.* [5] interpreted the magnetic state in terms of the heterogeneous coexistence of two phases - pure ferromagnetic (FM) and antiferromangetic (AFM) phases, while Yoshizawa *et al.* [6] consider the system to consist of a homogeneous canted antiferromagnetic (CAFM) phase. The homogeneous CAFM model is central for the interpretation of the extremely large CMR effect, because the electron-phonon and/or electron-magnon coupling, which results in the formation of Jahn-Teller polarons or magnetic polarons, play a central role [7, 8]. There is experimental evidence for this coupling at higher temperatures [9]. On the other hand, there is evidence that, as the temperature is decreased, the systems are not in a homogeneous single phase, but separated into two coexistent phases of FM+AF with mesoscopic length scale. Such observations render somewhat different interpretation of CMR effect based on more macroscopic view points. That is, the CMR materials tend to stabilize a "two-phase coexistence" state as a thermodynamically metastable state. Then, CMR effect is interpreted to be caused by the growth of FM-phase region, at an expense of AF-phase region [10, 30]. Indeed, some groups already observed nano-scale ferromagnetic structure (clusters, or even novel "red cabbage" or filaments) in $La_{1-x}Ca_xMnO_3$ and $Pr_{1-x}Ca_xMnO_3$ CMR systems by high resolution neutron diffraction or small-angle neutron scattering (SANS) [11~15, 26]. The understanding of the local and long-range magnetic and structural ordering in $Bi_{0.125}Ca_{0.875}MnO_3$ will help to develop an accurate model for CMR effect in certain systems.

Recent NMR results already showed evidence for the FM/AFM phase-separation in $Bi_{0.125}Ca_{0.875}MnO_3$ [16]. NMR is a powerful tool to reveal the spin configuration on the microscopic scale: nuclei in both FM and AFM domains will contribute to the $^{55}Mn$ NMR spectrum, and will produce two distinct resonance lines. Following this idea, Shimizu *et al.* carried out $^{55}Mn$ NMR measurements using a coherent spin-echo apparatus. NMR spectra were obtained by measuring the integrated spin-echo signal versus frequency. Two resonance peaks were observed: a low frequency line (~298MHz at H=0T) varies only slightly under different magnetic fields up to 4T indicating its AFM nature whereas the upper line (~318MHz at H=0T) shows strong dependence of external field with significant change both in peak position and shape which indicates its FM nature. This two-peak configuration is typical in similar systems such as $Ca_{1-x}Pr_xMnO_3$ (x<0.1), $La_{0.7}Ca_{0.3}MnO_3$ and $La_{0.7}Sr_{0.3}MnO_3$ [17, 18] and is discussed by most authors in terms of FM/AFM phase co-existence.

In this work we have utilized small angle neutron scattering (SANS), which is a powerful magnetic structure probe to detect the presence of microscopic FM clusters (with a scale from 1 nm to 100 nm) hosted in an AF background matrix. More detailed analysis of the data can give the size variation of FM clusters with varying temperatures and external magnetic fields.

## II. EXPERIMENT

We have synthesized polycrystalline $Bi_{0.125}Ca_{0.875}MnO_3$ sample by the standard ceramic technique. Stoichiometric mixtures of $Bi_2O_3$, $CaCO_3$ and $MnO_2$ were ground and pressed into pellets that were then calcined at 1000°C in air for 5 hours. Then the sample was reground and sintered at 1200°C in air for 5 hours. This was repeated 3 times. The synthesized sample was characterized by x-ray diffraction (XRD at the X14A of National Synchrotron Light Source of Brookhaven National Lab), Neutron Powder Diffraction (NPD, at the Chalk River Lab in Canada) and DC magnetization measurements (ZFC and FC performed with a Quantum Design SQUID magnetometer, MPMS-XL, between 5K and 300K). The multi-temperature XRD reveals lattice symmetry of Pnma (20K~300K, dominant lattice phase) and P21/m (20K~150K, minor lattice phase, less than 3%). The multi-temperature NPD reveals the mixing of G-type (20K~300K, dominant AF phase) and C-type AF long ordering (20K~150K, minor AF phase, less than 3%).

The SANS experiment was conducted at the SANS-2 facility of GKSS research center, Germany. Neutrons with a mean wavelength of λ=0.58nm (for detector distances of d=0.98m, 2.98m, 8.98m and 21.7m) and λ=1.16nm (for detector distance of d=21.7m) were selected by a rotating monochromator. Four different detector distances at d=0.98m, 2.98m, 8.98m and 21.7m were used in order to cover the full q spectrum ($0.01 nm^{-1}$ ~ $2.5 nm^{-1}$). The powder sample was centered in the sample chamber and then both temperature (T=20K, 100K, 108K, 120K, 150K, 200K) and magnetic field dependent (H=0T, 0.1T, 1T and 1.75T) measurements for the differential scattering cross section ($d\Sigma/d\Omega$) were performed. Data acquisition times were 30-60 minutes per measurement depending on the used distance. Scattered neutrons were recorded with an area detector of 256*256 pixels. The measured data were calibrated by the incoherent scattering of vanadium and corrected for the sample transmission, background scattering (from the sample holder) and detector response.

## III. RESULTS AND DISCUSSION

The total scattering consists of two parts, the nuclear scattering (originated from the pure structural background) as well as the magnetic scattering (originated from the spin structure, which is associated with the FM component). In order to separate the non-magnetic/nuclear background scattering from the magnetic part, magnetic fields H=0.1T, 1T and 1.75T are applied on the sample along the horizontal x direction. In this case, the magnetic scattering cross-section depends on the angle α between the scattering vector and the magnetic field. The total scattering cross-section can be written as:

$$\frac{d\Sigma}{d\Omega}(\vec{q}) = \frac{d\Sigma_n}{d\Omega}(\vec{q}) + \sin^2\alpha \frac{d\Sigma_m}{d\Omega}(\vec{q})$$

where

$$\alpha = \cos^{-1}\left(\frac{\vec{q}\cdot\vec{H}}{|\vec{q}||\vec{H}|}\right)$$

$\vec{q}$ is the scattering vector and $\vec{H}$ is the external magnetic field.

Thus the nuclear cross-section is measured at α=0° and 180°, while the sum of nuclear and magnetic cross-section is measured at α=90° and 270°. Practically the scattering curves will be calculated from about 10° wide sectors around those angles (α=0°, 90°, 180° and 270°). In this way any magnetic structure caused by the external field can be reflected by the anisotropy on the obtained SANS spectra, i.e., "nuclear+magnetic" scattering minus "nuclear" scattering (for similar cases, see [19~21]).

Utilizing the method above, we have obtained the magnetic SANS spectra for $Bi_{0.125}Ca_{0.875}MnO_3$. Fig. 2 shows an example (at magnetic field H=1.75T and T=20K). After careful examination of the spectra, we found they can be divided into 3 regions: (i) granular scattering region (q=0.01~0.06nm$^{-1}$); (ii) intermediate scattering region (q=0.06~0.7nm$^{-1}$); and (iii) cluster scattering region (q=0.7~2.4nm$^{-1}$).

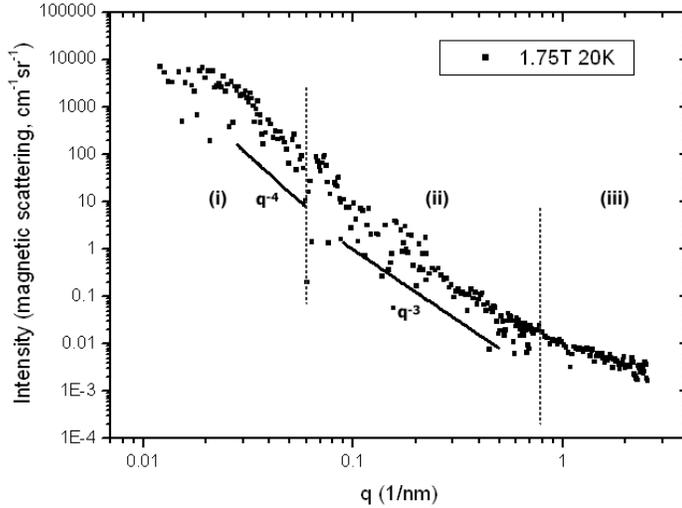

**Fig.2** An example magnetic SANS spectrum of $Bi_{0.125}Ca_{0.875}MnO_3$ (at H=1.75T and T=20K), which is obtained by subtracting the nuclear scattering part from the total (nuclear + magnetic) scattering intensity. The two solid lines (guide to eye) are power laws with exponents -4 and -3.

The granular scattering region is characterized by a Porod's $q^{-4}$ law feature occurs in the q range of 0.03~0.07 $nm^{-1}$ with a "head" (q=0.01~0.03 $nm^{-1}$) deviated from the Porod's $q^{-4}$ law. Since Porod's $q^{-4}$ law is a general rule for small-angle scattering of systems where smooth sphere particles are embedded in a matrix background, the SANS scattering curve in this region corresponds to a magnetic granular structure (with length scale of 30nm~100nm) in the sample [14]. It should be noted that the deviation of the "head" originates from the nonuniformity of granular size [20]. More data analysis reveals that this magnetic granular structure is not temperature-dependent, which indicates it is not relevant to the FM component found below Tc.

In the intermediate scattering region, the q dependence is represented by a $q^{-\alpha}$ law ($\alpha$=2.4~3.4). This $q^{-\alpha}$ profile is attributed to the scattering from some "non-spherical" or anisotropic ferromagnetic objects in the system (the detail will be discussed later in this paper). Data analysis indicates this FM configuration are temperature-dependent, especially in the higher q range of 0.2~0.7 $nm^{-1}$ (length scale of 2~10nm). These FM objects contribute to the spontaneous magnetization below Tc.

The most significant field and temperature dependent anisotropy occurs in the cluster scattering region, shown in Fig.3 (a) ~ (c). It should be noted in these figures: 1) at T<Tc, when the field is increased, the magnetic scattering will increase, which implies the FM component (1~3nm in length scale) grows with the applied external field; 2) at T>Tc, there is no apparent magnetic scattering, which implies no, or only a little bit of, FM component exists in this length scale at T>Tc. This systematic trend is consistent with the ZFC/FC magnetization measurements (see Fig.4 (b) below).

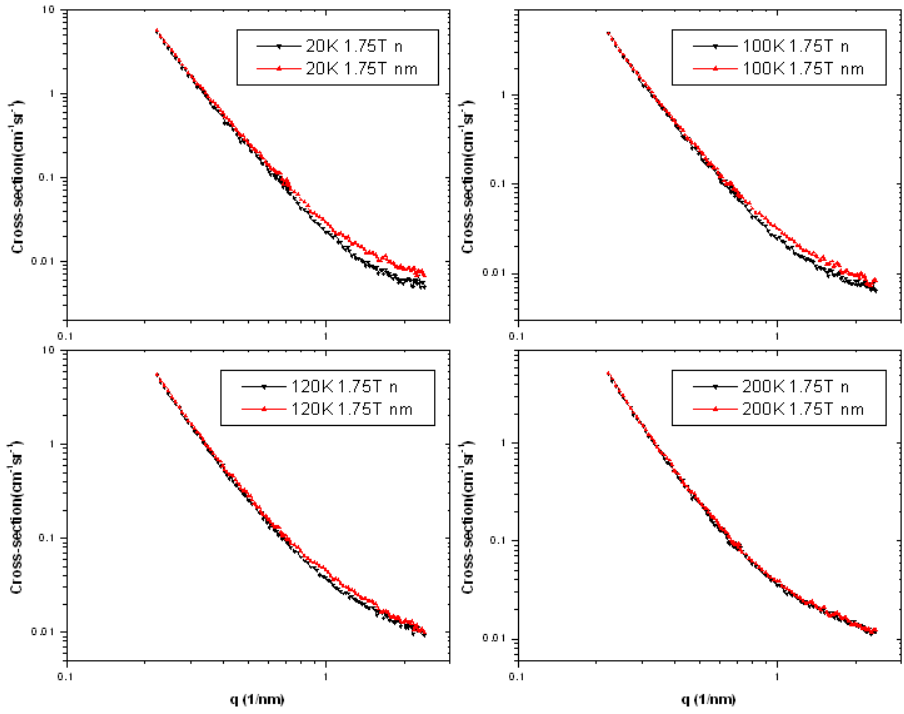

**Fig.3**   (a) Magnetic cross-section at H=1.75T from SANS of $Bi_{0.125}Ca_{0.875}MnO_3$.

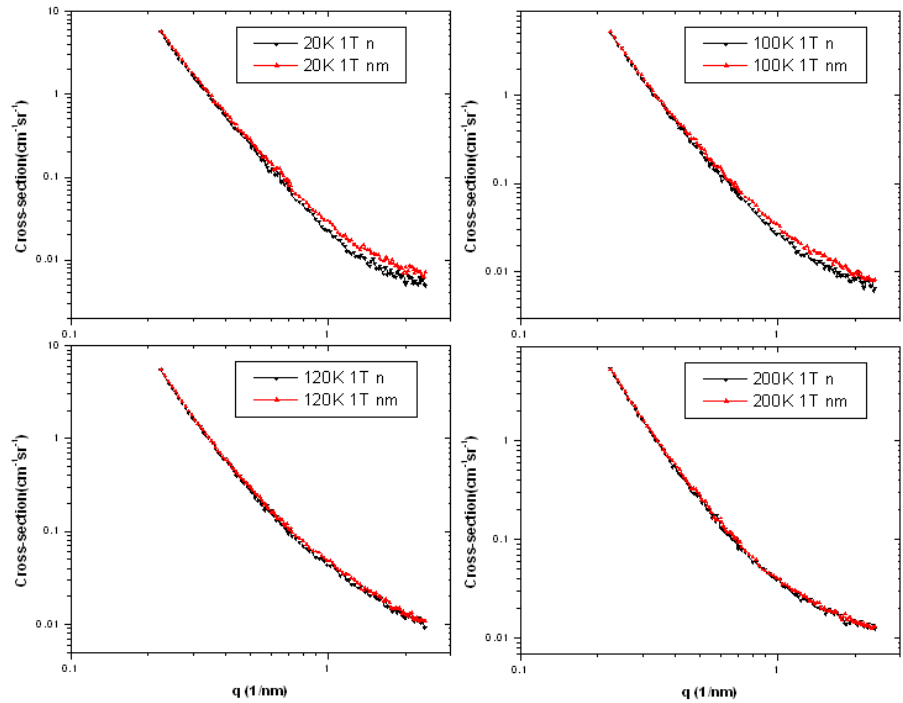

**Fig.3**   (b) Magnetic cross-section at H=1T from SANS of $Bi_{0.125}Ca_{0.875}MnO_3$.

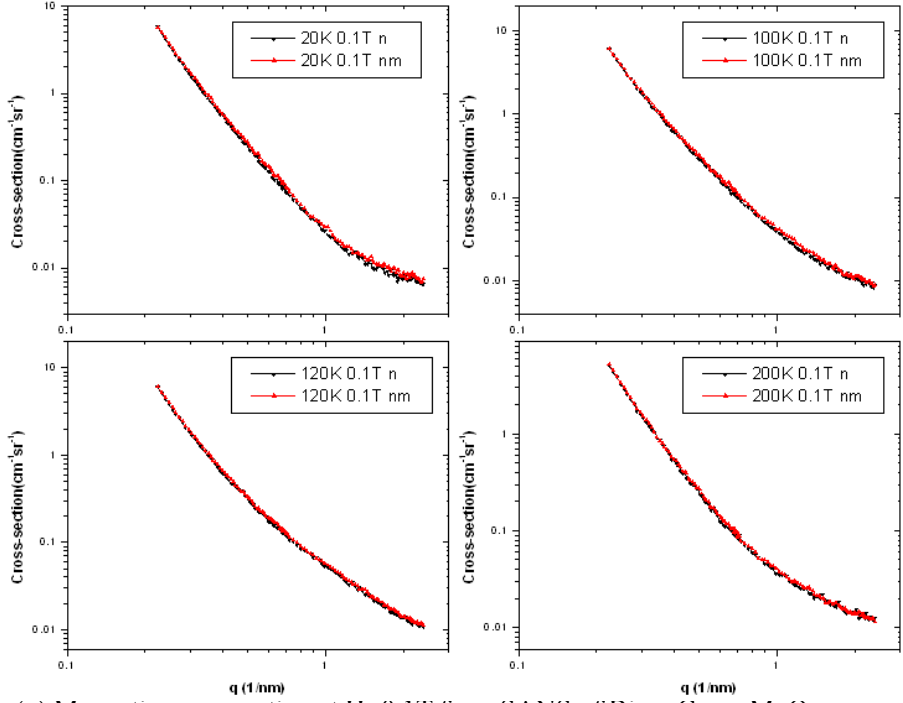

**Fig.3** (c) Magnetic cross-section at H=0.1T from SANS of $Bi_{0.125}Ca_{0.875}MnO_3$.

No significant temperature-dependent anisotropy has been found in the low-q range (0.01~0.2 $nm^{-1}$), which indicates the spin structure associated with the FM component below Tc only exists in the scale of 1~10 nm.

In order to extract more detailed information regarding the FM component, the q dependence of the magnetic scattering at q range of 0.8~2.6$nm^{-1}$ has been generally modeled by the Lorentzian profile, which describes a clustering structure [22-29], as:

$$\frac{d\Sigma_m}{d\Omega}(q) = \frac{A}{q^2 + \kappa^2}$$

where A is the scattering amplitude and $\kappa=1/\xi$, $\xi$ is the magnetic correlation length of the FM clusters (its average size). By fitting of the scattering curves with the Lorentzian profile, the scattering amplitude A (represents the volume fraction of the FM clusters in the whole system) and magnetic correlation length $\xi$ (represents the cluster size) can be extracted.

Fig.4 (a) shows the fitted results for scattering amplitude A under different temperatures and fields. It can be found that: 1) the magnetic scattering amplitudes will increase when temperature is decreased across Tc. 2) decreasing of the external field will decrease the magnetic scattering amplitudes. In order to illuminate the physical meaning behind this phenomenon, the results above are compared with the magnetization measurements done by Woo et al. [4], shown in Fig. 4 (b). The systematic consistency between these two independent measurements indicate that the FM component below Tc in $Bi_{0.125}Ca_{0.875}MnO_3$ originates from the FM clustering structure in 1~10nm.

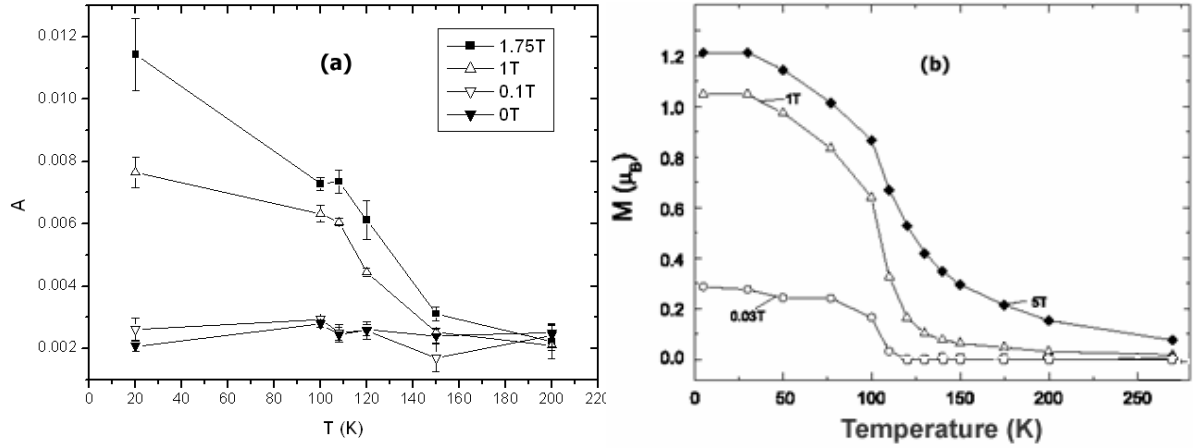

**Fig.4** (a) Fitted scattering amplitudes (A) for magnetic SANS scattering curves at the q range of 0.8~2.6 $nm^{-1}$, modeled by the Lorentzian profile. (b) Temperature dependence of the magnetization induced by magnetic fields [4].

The fitting results for A, $\kappa$ ($1/\xi$) and $\xi$ under different temperatures and fields are listed in Tab.1 (a) ~ (d). It should be noted that: for data sets of T<Tc and external field H=1T and 1.75T, the $\kappa$ can be easily obtained through the fitting process. For T>Tc or weak field (H=0T and 0.1T), the obtained are uncertain due to the weak signal at high temperature or low field. However, they generally maintain values less than $0.3nm^{-1}$, which suggests that at T>Tc or zero field, bigger (but very few) FM clusters (size > 3nm) already exist.

**Tab.1 (a)** Fitting results for A, $\kappa$ and $\xi$ at H=1.75T.

| Temperature(K) | A | $\kappa$ | $\xi$ (nm) |
| --- | --- | --- | --- |
| 20  | 0.0114 ± 0.002  | 0.91 ± 0.24 | 1.10 ± 0.28 |
| 100 | 0.0073 ± 0.0002 | 0.45 ± 0.03 | 2.22 ± 0.15 |
| 108 | 0.0073 ± 0.0004 | 0.28 ± 0.07 | 3.57 ± 0.89 |
| 120 | 0.0061 ± 0.0006 | 0.13 ± 0.14 | - |
| 150 | 0.0031 ± 0.0002 | 0.05 ± 0.05 | - |
| 200 | 0.0022 ± 0.0005 | 0.28 ± 0.31 | - |

**Tab.1 (b)** Fitting results for A, κ and ξ at H=1T.

| Temperature(K) | A | κ | ξ (nm) |
|---|---|---|---|
| 20 | 0.0077 ± 0.0004 | 0.59 ± 0.07 | 1.69 ± 0.20 |
| 100 | 0.0063 ± 0.0003 | 0.30 ± 0.05 | 3.33 ± 0.56 |
| 108 | 0.0061 ± 0.0001 | 0.01 ± 0.01 | - |
| 120 | 0.0045 ± 0.0001 | 0.03 ± 0.04 | - |
| 150 | 0.0025 ± 0.0001 | 0.07 ± 0.08 | - |
| 200 | 0.0021 ± 0.0002 | 0.01 ± 0.01 | - |

**Tab.1 (c)** Fitting results for A, κ and ξ at H=0.1T.

| Temperature(K) | A | κ | ξ (nm) |
|---|---|---|---|
| 20 | 0.0026 ± 0.0004 | 0.03 ± 0.06 | - |
| 100 | 0.0029 ± 0.0001 | 0.02 ± 0.03 | - |
| 108 | 0.0025 ± 0.0003 | 0.01 ± 0.01 | - |
| 120 | 0.0026 ± 0.0003 | 0.07 ± 0.13 | - |
| 150 | 0.0017 ± 0.0005 | 0.35 ± 0.33 | - |
| 200 | 0.0024 ± 0.0004 | 0.01 ± 0.01 | - |

**Tab.1 (d)** Fitting results for A, κ and ξ at H=0T.

| Temperature(K) | A | κ | ξ (nm) |
|---|---|---|---|
| 20 | 0.0021 ± 0.0002 | 0.12 ± 0.13 | - |
| 100 | 0.0028 ± 0.0001 | 0.10 ± 0.22 | - |
| 108 | 0.0024 ± 0.0002 | 0.01 ± 0.01 | - |
| 120 | 0.0026 ± 0.0002 | 0.11 ± 0.14 | - |
| 150 | 0.0024 ± 0.0002 | 0.01 ± 0.01 | - |
| 200 | 0.0025 ± 0.0002 | 0.02 ± 0.03 | - |

Some groups have attributed the FM clusters in similar systems ($La_{1-x}Ca_xMnO_3$ and $Pr_{1-x}Ca_xMnO_3$) to the presence of magnetic polarons [26]. However, in those systems, when temperature is decreased to Tc, the FM clusters will grow in size and merge into a connected FM network. This is different from what we have found in $Bi_{0.125}Ca_{0.875}MnO_3$, where, when temperature is reduced below Tc, or external field is increased, the average cluster size will decrease. This contradicts the common view that lowering temperature or increasing of field will cause the clusters to grow in size. The major question raised here is why ξ drops while the overall magnetization (described by A) is increased with the reduced temperature and enhanced field. Our proposed model is, the common view ignores the effects from cluster number growth below Tc or in field - more small clusters could compensate the reduction of cluster size and increase the overall magnetization.

The fitting results (with the $I=Aq^{-\alpha}$ profile) for the SANS spectra in $q=0.2\sim0.7 nm^{-1}$ give additional information, which is shown in Fig. 5. The behavior of the scattering amplitude A (volume fraction of the scattering objects in the system) mimics the magnetization measurement (see Fig. 4 (b)) very well. It is already known that the power-law exponent $\alpha$ reflects the spatial symmetry of the scattering objects. While Porod's $q^{-4}$ law indicating round-shape (spherical) scattering objects, exponent $\alpha$ smaller than 4 but larger than 2 demonstrates the existence of rod-like or flat-shape particles [31]. These non-spherical FM objects (in length scale of 2 ~ 10nm) may be originated from the merger of two or more smaller FM clusters.

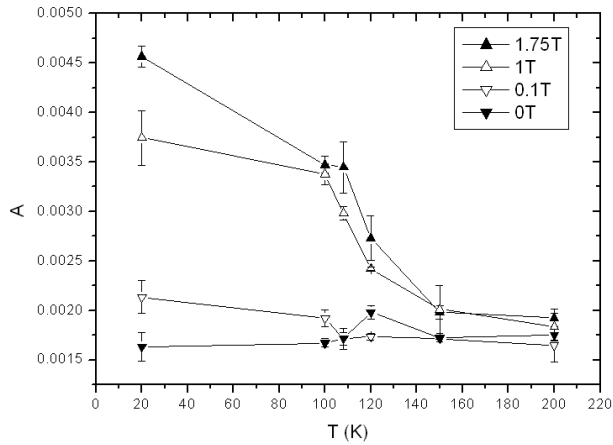

**Fig.5** The fitting results of the temperature-dependent scattering amplitude A of $Bi_{0.125}Ca_{0.875}MnO_3$ in the "intermediate scattering region" ($\alpha$ =2.5~3.4). They mimic the magnetization measurements in Fig.4 (b) very well.

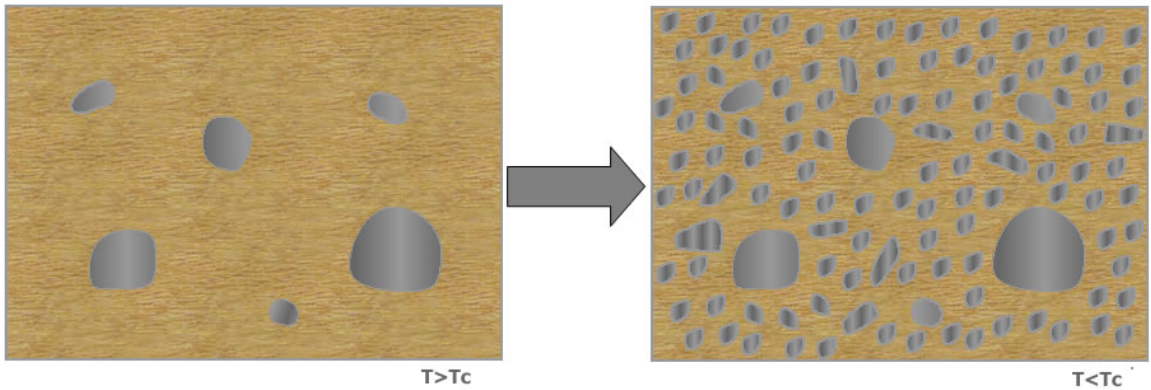

**Fig.6** Diagram of the proposed model for the FM-cluster number-growth below Tc.

In order to give a clearer picture of the system in the temperature region below Tc or with an applied external field, we sketch our proposed model in Fig.6. Above Tc large magnetic clusters exist. When temperature is decreased below Tc or the external field is increased, more and smaller FM clusters (in length scale of 1~3nm) will emerge, causing a decreased average cluster size but an increased overall magnetization. Some of these FM clusters may merge into larger rod-like or flat-shape objects (2~10nm), which give the $q^-$ scattering in the intermediate scattering region.

The transportation property of $Bi_{0.125}Ca_{0.875}MnO_3$ observed by Chiba *et al*. [1] sheds more light on our model. The increase in resistivity observed at low temperature in this system is characteristic of what would be expected for a mixed phase metallic-insulating system with percolation driven transport [32]. The details of the transport are linked to the connectivity between the metallic regions. The transportation property of $Bi_{0.125}Ca_{0.875}MnO_3$ indicates the connectivity between the FM clusters is quite low, which is different from the FM component in $(La,Pr,Ca)MnO_3$ [30] and $(Pr,Y,Ca)MnO_3$ [26] systems.

The low connectivity of the clusters also affects the magnetic properties. The isolation of FM clusters can account for the low magnetization in $Bi_{0.125}Ca_{0.875}MnO_3$: the AFM background occupies such a large fraction of the whole system that the magnetization at low temperature can never get saturated to the theoretical limit of 3.1 Bohr magnetons per Mn site.

## IV. CONCLUSION

In conclusion, by conducting temperature and magnetic field dependent small-angle neutron scattering measurements on $Bi_{0.125}Ca_{0.875}MnO_3$, we have identified ferromagnetic spin clusters embedded in the antiferromagnetic matrix at low temperature. With a reduction in temperature or the application of external magnetic field, the clusters begin to form in larger numbers and smaller average size.

We gratefully acknowledge Dieter Lott for kind assistance in small angle neutron scattering (SANS) measurements at GKSS and Dr. Guerman Popov and Prof. Martha Greenblatt (Rutgers University) for SQUID magnetization measurements. This project is supported by National Science Foundation DMR-0209243, DMR-0512196 and NSF INT-0233316.